\documentclass[trackchanges,twocolumn]{aastex701}

\usepackage{amsmath}
\usepackage{bm}
\usepackage{mathtools}
\usepackage{xspace}
\usepackage{xcolor}

\newcommand{\Rstar}{\ensuremath{R_{\star}}\xspace}

\newcommand{\rhostar}{\ensuremath{\rho_{\star}}\xspace}

\newcommand{\Rp}{\ensuremath{R_p}\xspace}

\newcommand{\Me}{\ensuremath{M_{\oplus}}\xspace} 
\renewcommand{\Re}{\ensuremath{R_{\oplus}}\xspace}

\newcommand{\Rsun}{\ensuremath{R_{\odot}}\xspace }

\newcommand{\K}{\ensuremath{\mathrm{K}}\xspace}

\newcommand{\numplanets}{1646\xspace}
\newcommand{\numstars}{1209\xspace}

\newcommand{\alderaan}{{\tt ALDERAAN}\xspace}
\newcommand{\e}{{\langle}e{\rangle}}

\begin{document}

\title{No strong associations between eccentricity and orbital architecture in Kepler compact multis}

\author[orcid=0000-0003-0742-1660,sname=Gilbert]{Gregory J. Gilbert}
\affiliation{Department of Astronomy, California Institute of Technology, Pasadena, CA 91125, USA}
\affiliation{Department of Physics \& Astronomy, University of California Los Angeles, Los Angeles, CA 90095, USA}
\email[show]{ggilbert@caltech.edu}  

\author[orcid=0000-0003-0967-2893,sname=Petigura]{Erik A. Petigura} 
\affiliation{Department of Physics \& Astronomy, University of California Los Angeles, Los Angeles, CA 90095, USA}
\email{petigura@astro.ucla.edu}  

\author[orcid=0009-0008-6123-1634,sname=Entrican]{Paige M. Entrican} 
\affiliation{Department of Physics \& Astronomy, University of California Los Angeles, Los Angeles, CA 90095, USA}
\email{pentrican10@g.ucla.edu}  

\begin{abstract}

The dynamical history of a planetary system is recorded in the present day architecture of its constituent planets' sizes, orbital periods, and eccentricities. Studying the relationships between these quantities for large populations provides a window into the processes by which planetary systems form and evolve. Recently, \citet{Gilbert2025} performed a hierarchical Bayesian analysis of \numplanets planets from the Kepler census, demonstrating a strong relationship between planet radius $R_p$ and orbital eccentricity $e$. Here, we build upon that work to search for correlations between eccentricity and system architecture, focusing on compact systems of small planets. We find that small planets on short orbits ($P < 4$ days) show evidence of tidal circularization. This trend is well established for Jovian planets but a novel finding for super-Earths and sub-Neptunes. We reproduce the known wherein trend single-transiting systems posses elevated eccentricities relative to their multi-transiting counterparts. We further show that systems with two transiting planets have higher eccentricities than those with three or more transiting planets. When compared to population synthesis models, these multiplicity-eccentricity relationships imply that Kepler singles have intrinsic multiplicity ${\sim}3$ and Kepler multis have intrinsic multiplicity ${\sim}4{-}6$. We detect no statistically significant associations between eccentricity and planetary period ratios, gap complexity, size inequality, or size ordering. We interpret these findings as evidence either in favor of a quiescent formation history or against dynamical processes which excite eccentricity but not inclination. Sub-significant relationships between eccentricity and architecture imply that subtle, multi-factor trends may be detectable in the future using more sophisticated statistical techniques.

\end{abstract}


\section{Introduction}\label{sec:intro}

The dynamical histories of planetary systems are recorded in their present day arrangements of planet compositions, sizes, and orbits. One relatively direct probe of dynamical history is orbital eccentricity. Some dynamical processes tend to excite eccentricity (e.g., planet-planet scattering, \citealt{Chatterjee2008, FordRasio2008, JuricTremaine2008}), while others damp it (e.g., tidal friction, \citealt{Rasio1996, Jackson2008}; disk migration, \citealt{TanakaWard2004, PapaloizouTerquem2006, KleyNelson2012}). A less direct but equally powerful probe of dynamical history is system size and spacing architecture (i.e. the arrangement of planet sizes and orbital periods), because dynamical instabilities tend to disrupt orderly size and spacing architecture \citep[e.g.,][]{PuWu2015, Izidoro2017, Lambrechts2019}, whereas more quiescent modes of formation and migration tend to preserve more regular architectures \citep[e.g.,][]{IdaLin2008, IdaLin2010, Adams2020}. These processes are stochastic, so any confident inferences regarding the relative importance of different formation channels must by made through demographic analysis of large populations of planets and systems.

Several robust trends have emerged regarding the size, spacing, and eccentricity architectures of planetary systems. Small planets ($R_p < 3.5~\Re$) are abundant (${\sim}1$ planet per star with $P < 100$ days), tend to have low eccentricities $\e \approx 0.05$, and are often found in ``peas-in-a-pod'' systems, sharing roughly uniform sizes and log-uniform spacing with their siblings \citep{Petigura2013, Millholland2017, Weiss2018, VanEylen2019, Gilbert2025}. In contrast, large planets ($\Rp > 3.5~\Re$) are an order of magnitude more rare, possess elevated eccentricities $\e\approx 0.2$, and are more often found in single-planet systems \citep{Howard2010, Steffen2012, Knutson2014, Kipping2013-eccentricity, VanEylen2019, Gilbert2025}. Taken together, these trends imply a paradigm whereby the influence of dynamical history on system architecture scales with planet size.

Further clues into the planet formation process can be gleaned by considering multiple demographics trends in concert. Because a large sample size is vital for robust multi-dimensional inference, the majority of these demographics analyses have been performed using the Kepler census \citep{Borucki2010, Thompson2018}. In a recent analysis, \citet{Gilbert2025} considered a sample of \numplanets uniformly characterized Kepler planets, finding a sharp transition from low to high mean eccentricity at $R_p \approx 3.5~\Re$, coinciding with transitions in planet occurrence rate \citep{Fulton2017} and host star metallicity \citep{Buchhave2012}, which they interpret as evidence that giant planets play a dominant role in mediating dynamical interactions. 

Here, we build upon the work of \citet{Gilbert2025} to study how orbital eccentricity varies as a function of system architecture. The aim of this paper is to explore whether correlations exist between various probes of dynamical history for Kepler compact multi-planet systems (hereafter ``compact multis''), i.e. systems of multiply-transiting small planets ($R_p \leq 3.5~\Re$) inside $P < 100$ days.



This paper is organized as follows. In \S\ref{sec:methods} we review our data sample and analysis methods. In \S3 -- \S6 we search for relationships between eccentricity and the presence of giant planets (\S\ref{sec:giants}), observed transit multiplicity (\S\ref{sec:multiplicity}), system spacing architecture (\S\ref{sec:spacing}), system size architecture (\S\ref{sec:size}), and dynamical temperature (\S\ref{sec:dyntemp}). In \S\ref{sec:discussion} we discuss the implications of these relationships, and in \S\ref{sec:conclusion} we summarize our findings and our main conclusions.

\section{Data and Methods}\label{sec:methods}

For this analysis, we adopted the same data and methods as those used in \citet{Gilbert2025}. A brief overview is described below.

\subsection{Planet sample}

Our sample consisted of \numplanets planets with orbital periods $P = 1{-}100$ days and sizes $\Rp = 0.5{-}16~\Re$.  All planets were uniformly detected and vetted by the final \textit{Kepler} project data release \citep[DR25;][]{Thompson2018}. The planets orbit \numstars main-sequence stars of spectral types F, G, and K ($R_{\star} = 0.7{-}1.4$~\Rsun, effective temperature $T_{\rm eff} = 4700{-}6500$~K, and surface gravity $\log g > 4.0$). Stellar input properties were taken from the \textit{Gaia}--\textit{Kepler} stellar catalog \citealt{GaiaDR2, Berger2020}. To ensure robust measurement of planet and host star properties, we filtered the initial input catalog to exclude stars with a flux contamination fraction greater than $5\%$ \citep{Furlan2017} or a Gaia Renormalized Unit Weight Error $\text{RUWE} \geq 1.4$, as such stars have high probability of being unresolved binaries \citep{Wood2021}. Finally, we removed all stars with $\sigma(\Rstar)/\Rstar > 20\%$.

\subsection{Extracting eccentricity from transit photometry}

To derive planet properties, we fit \textit{Kepler} transit photometry for all \numplanets planets using the \alderaan transit-fitting pipeline\footnote{\url{https://www.github.com/gjgilbert/alderaan}}. \alderaan ingests pre-search data conditioning simple aperature photometry \citep[PDCSAP;][]{Stumpe2012}, removes stellar and instrumental variability using Gaussian process regression \citep{ForemanMackey2017-celerite}, and infers a regularized model for any dynamical transit timing variations (TTVs) using cross-correlation template matching \citep{Mazeh2013, Holczer2016}. \alderaan then produces posterior samples of orbital period $P$, planet-to-star radius ratio $\Rp/\Rstar$, transit impact parameter $b$, and first-to-fourth contact transit duration $T_{14}$ using dynamic nested sampling \citep{Skilling2004, Skilling2006, Higson2019, Speagle2020} coupled to an analytic transit model \citep{MandelAgol2002, Kreidberg2015}. The transit model assumes a quadratic stellar limb darkening profile \citep{Kipping2013-limbdarkening} and applies modestly informative priors calculated from Gaia-derived stellar parameters \citep{Husser2013, Parviainen2015, GaiaDR2, Berger2020}. The model specification was developed to provide unbiased estimates of $\Rp/\Rstar$, $b$, and $T_{14}$ to better than ${\sim}5\%$ precision while preserving a record of subtle parameter covariances \citep{Carter2008, Gilbert2022-pseudodensity}. \alderaan has been rigorously tested via a suite of injection-and-recovery tests and has been demonstrated to reliably recover transit properties.

To convert posterior samples of $\{P, \Rp/\Rstar, b, T_{14}\}$ to samples of eccentricity $e$ and argument of periastron $\omega$, we applied the importance sampling scheme developed by \citet{MacDougall2023}. This method leverages the so-called photo-eccentric effect \citep{Ford2008, DawsonJohnson2012, Kipping2014-asterodensity} by combining \alderaan-derived posterior samples $\{P, \Rp/\Rstar, b, T_{14}\}$ with independent measurements of stellar density $\rhostar$ \citep{GaiaDR2, Berger2020} to produce samples of $\{P, \Rp/\Rstar, b, e, \omega, \rhostar\}$. These samples are equivalent to those that would have been produced had the transits been fit using an eccentric Keplerian model, but with much lower computational overhead.

The \texttt{ALDERAAN} pipeline produces posteriors on individual planet $\Rp/\Rstar$ and $T_{14}$ which are consistent to within ${\sim}5\%$ of values produced by the final Kepler data release \citep[DR25,][]{Thompson2018}, save for the few targets with spurious results in DR25 (for example, from mischaracterized grazing transits); posteriors on $P$ and $t_0$ are nearly identical between the two catalogs because orbital period is exquisitely constrained by transit photometery. The most critical improvement of \texttt{ALDERAAN} over DR25 for this work is that \texttt{ALDERAAN} produces accurate and complete posteriors of $b$ and preserves covariances between $b$, $T_{14}$, and $e$, which are needed to correctly infer the eccentricity distribution.

\subsection{Inferring the eccentricity distribution}

To infer the eccentricity distribution, we implement an approximate hierarchical Bayesian framework \citep{Hogg2010}. This method is rapidly becoming the standard method used to infer exoplanet eccentricities \citep[e.g.,][]{VanEylen2019, Bowler2020, SagearBallard2023, Gilbert2025, Sagear2025}.

Following the \citet{Hogg2010} formalism, the likelihood $\mathcal{L}_\alpha$ for the population distribution $f(e)$ is
\begin{equation}\label{eq:hbayes-likelihood-with-detection-bias}
    \mathcal{L}_{\alpha} = \prod_{n=1}^N \frac{1}{K} \sum_{k=1}^K \frac{f_\alpha(e_{nk})}{p_0(e_{nk})} \left( \frac{1-e_{nk}^2}{1+e_{nk}\sin\omega_{nk}} \right)
\end{equation}
where $e_{nk}$ are chains of $K$ eccentricity samples for each of the $N$ planets in the sample. The term $p_0(e)$ is the uninformative ``interim'' prior on $e$ applied during transit modeling and the term $f_\alpha(e)$ is an informative ``updated'' population distribution we wish to infer; the subscript $\alpha$ denotes the vector of hyper-parameters describing $f(e)$. Conceptually, the ratio $f_\alpha/p_0$ captures the degree to which our informative eccentricity population model improves upon an uninformative prior, marginalized over all covariant quantities. For our particular application to transiting planets, the right-hand $\{e_{nk},\omega_{nk}\}$ term in parentheses corrects for a geometric biases which leads to preferential detection of exoplanets on eccentric orbits compared to circular orbits \citep{Barnes2007, Burke2008, Kipping2014-ecc-priors}.

Following \citet{Gilbert2025}, we adopted a regularized histogram as the functional form of the eccentricity distribution \citep[see also][]{ForemanMackey2014-occurrence, Masuda2022}. For this specific application, we use 25 histogram bins, which enter Equation \ref{eq:hbayes-likelihood-with-detection-bias} as logarithmic bin heights $\alpha$. Regularization (smoothing) is enforced via a Gaussian process prior which imposes correlation between adjacent bins. The advantage of this model is that it is highly flexible and makes no assumption about the form of the eccentricity distribution other than that it must be smooth and continuous. After fitting the full population of \numplanets planets, we ``froze'' the distribution to produce a parent template. This parent distribution is quasi-exponential in morphology, with a maximum at $e=0$ and a monotonic decline toward zero at $e=1$. \citet{Gilbert2025} found that the shape of the eccentricity distribution is self-similar across a wide range of planet sizes ($0.5 \leq \Rp/\Re \leq 16$), modulo a scale factor, so it it permissible to adopt the same parent distribution for all subpopulations derived from our \numplanets sample.

To measure differences in the mean, variance, and kurtosis of the eccentricity distribution for various sub-populations of planets, we perturbed the empirical parent distribution $f_0$ as
\begin{equation}\label{eq:empirical-pdf}
    \ln f_E(e;\nu,h) = \nu \left(\ln f_0(he)\right) + (1-\nu) \left(\ln f_0(0)\right)
\end{equation}
where $\nu$ and $h$ are positive coefficients controlling the scaling in the vertical and horizontal directions, respectively. Following transformation, the resultant $f_E$ was numerically normalized. To first order, $\nu$ sets the heaviness of the distribution tail (by scaling logarithmically in the vertical direction) and $h$ sets the width of the distribution core (by scaling linearly in the horizontal direction).

In the results that follow (\S3--6) we report both the full derived sub-population eccentricity distributions $f(e)$ as well as the mean eccentricities $\e$ of these distributions as a summary statistic.

\section{Giant Planets}\label{sec:giants}

Large planets are known to possess elevated eccentricities relative to their smaller counterparts \citep{Kipping2013-eccentricity, VanEylen2019, Gilbert2025}. Recently, \citet{Gilbert2025} demonstrated that a sharp transition in mean eccentricity $\e$ occurs at threshold radius $R_p \approx 3.5\Re$, wherein smaller planets have $\e = 0.05$ and larger planets have $\e = 0.20$. Motivated by this discovery, in this work we defined ``small planets'' as any objects with sizes between $0.5 \leq \Rp/\Re \leq 3.5$ and ``giant planets'' as any objects with sizes between $3.5 < \Rp/\Re \leq 16$.

To assess the role of giant planets in mediating orbital eccentricities, we split the small-planet population into two groups: (1) small planets in multi-planet systems \textit{with} known transiting giant planet companions, and (2) small planets in multi-planet systems \textit{without} known transiting giant companions. To avoid introducing biases and to treat the sample uniformly, we did not consider non-transiting giants (e.g. those detected by Doppler surveys), even if such planets are known to exist. We further limited this comparison to multi-planet systems in order to control for the known discrepancy between eccentricities of planets in single- vs multi-transiting systems \citep{VanEylen2019, Gilbert2025}. For each group, we inferred the eccentricity distribution $f(e)$ and mean eccentricity $\e$ following the procedure described in \S\ref{sec:methods}.

\begin{figure*}
    \centering
    \includegraphics[width=0.95\linewidth]{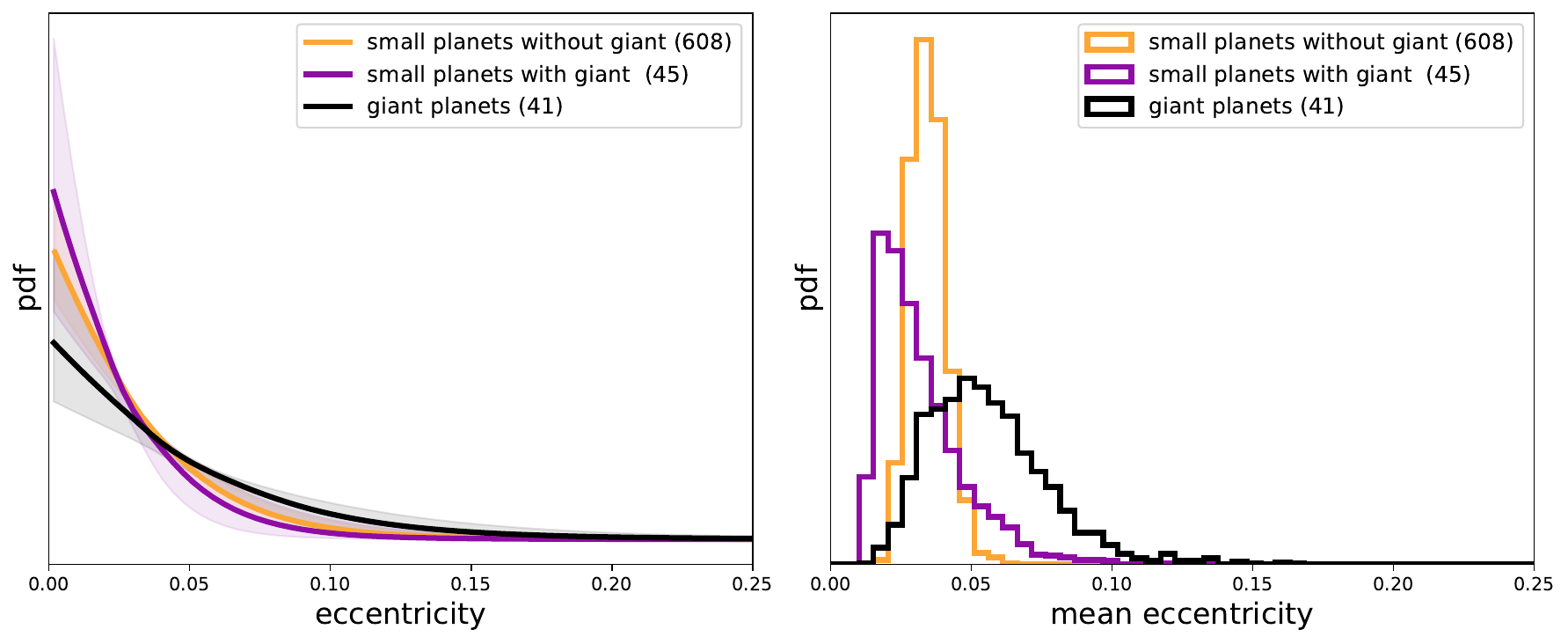}
    \caption{The eccentricity of small planets in multi-planet systems for planets with vs without giant companions. The number of planets in each sub-population is indicated in parentheses in the figure legend. \textit{Left panel}: Sub-population distributions $f(e)$. Dark solid lines indicate median retrieved distribution, and shaded regions indicate the 16th-84th percentile confidence interval. \textit{Right panel}: Mean eccentricity $\e$ for each of the sub-populations.}
    \label{fig:ecc-giants}
\end{figure*}

We found that the mean eccentricity of small planets in systems with giants, $\e = 0.028^{+0.018}_{-0.010}$,  is consistent with that of small planets in multi-planet systems without giants, $\e = 0.034^{+0.007}_{-0.006}$, to within $1\sigma$ (Figure \ref{fig:ecc-giants}). These values are somewhat lower than the mean eccentricity of the giant planets themselves, $\e = 0.053^{+0.022}_{-0.018}$, although even these agree within $1\sigma$ as well.

Even though we found no strong correlation between $\e$ and the presence of giant planets, we nevertheless excluded all systems hosting giant planets from the remainder of our analyses. Only 45 out of 672 planets (${\sim}7\%$) in the sample were excluded by this criterion, so the impact on statistical power was negligible.

\section{Multiplicity}\label{sec:multiplicity}

To explore the relationship between transit multiplicity and eccentricity, we split the small planets into four sub-populations, $N=1$ (i.e., single-transiting systems), $N=2$, $N=3$, and $N\geq4$, where N is the number of detected transiting planets between $P=1-100$ days and $0.5 \leq R_p/\Re < 3.5$. For each sub-population, we then inferred the eccentricity distribution $f(e)$ and mean eccentricity $\e$ following the procedure described in \S\ref{sec:methods}.

\begin{figure*}
    \centering
    \includegraphics[width=0.95\linewidth]{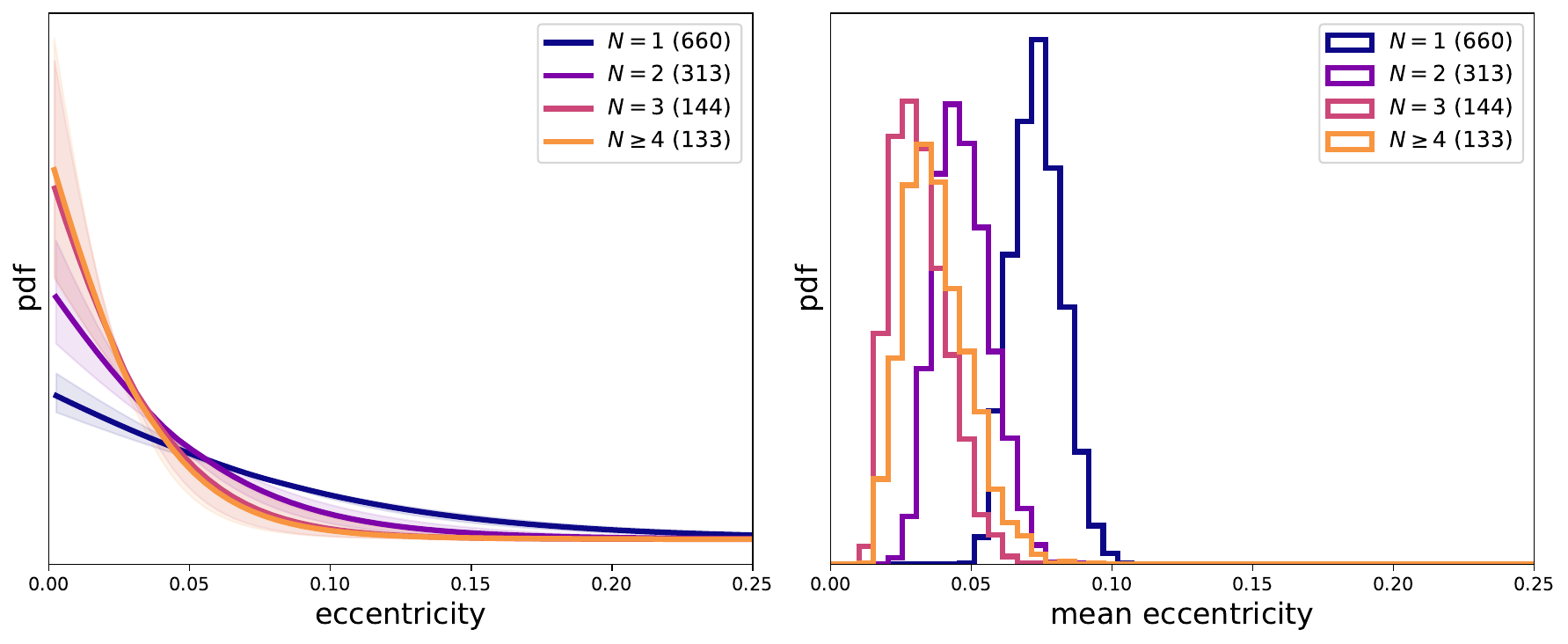}
    \caption{The eccentricity of small planets as a function of observed multiplicity. The number of planets in each sub-population is indicated in parentheses in the figure legend. \textit{Left panel}: Sub-population distributions $f(e)$. Dark solid lines indicate median retrieved distribution, and shaded regions indicate the 16th-84th percentile confidence interval. \textit{Right panel}: Mean eccentricity $\e$ for each of the sub-populations.}
    \label{fig:ecc-multiplicity}
\end{figure*}

We measured $\e_1 = 0.073 \pm 0.009$, $\e_2 = 0.046 \pm 0.009$, $\e_3 = 0.031 \pm 0.010$, $\e_{4+} = 0.036 \pm 0.012$ for planets in single, double, triple, and high multiplicity ($N \geq 4)$ systems, respectively (Figure \ref{fig:ecc-multiplicity}). These values reproduce the established trend that single-transiting systems have a higher mean eccentricity compared to multi-transiting systems \citep{VanEylen2019, Gilbert2025}. The mean eccentricities of $N=3$ and $N=4+$ systems were indistinguishable, while $\e$ for doubles fell between $\e$ for singles and $\e$ for systems with three or more planets.

\section{Spacing architecture}\label{sec:spacing}

\subsection{Orbital period}

To explore the relationship between eccentricity and orbital period, we split the small planet sample into six log-uniformly spaced bins between $P = 1-64$ days, treating single-transiting and multi-transiting systems independently.

\begin{figure}
    \centering
    \includegraphics[width=0.95\linewidth]{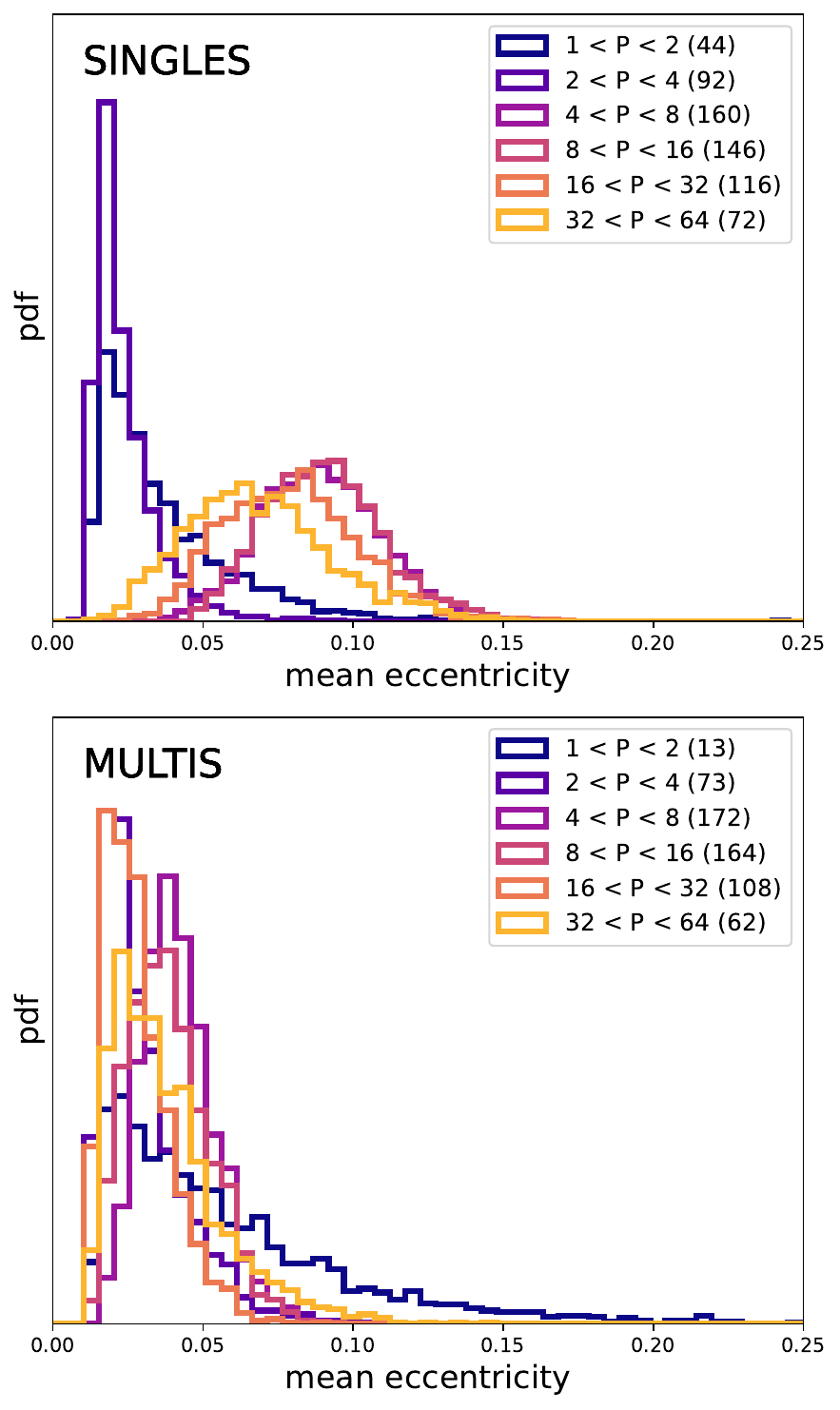}
    \caption{The eccentricity of small planets as a function of orbital period, split between singles (top panel) and multis (bottom panel). The number of planets in each sub-population is indicated in parentheses in the figure legend. Planets in single-transiting systems with $P < 4$ days show evidence of circularization, and planets in multi-transiting systems at all periods have on average low-$e$ orbits.}
    \label{fig:ecc-period}
\end{figure}

We found that inside $P < 4$ days, planets in singles have mean eccentricity $\e \approx 0.03^{+0.02}_{-0.01}$, whereas planets outside $P > 4$ days have $\e \approx 0.08^{+0.02}_{-0.02}$. These values are consistent with tidal circularization for planets on short-period orbits \citep{Rasio1996, Jackson2008}. Planets in multi-transiting systems all have low mean eccentricity $\e \approx 0.03^{+0.02}_{-0.01}$ across the full range periods we considered ($P = 1-64$ days); inferred $\e$ in each bin was consistent with all others to within $1\sigma$ (Figure \ref{fig:ecc-period}).

\subsection{Two-planet systems: period ratios}

\begin{figure}
    \centering
    \includegraphics[width=0.95\linewidth]{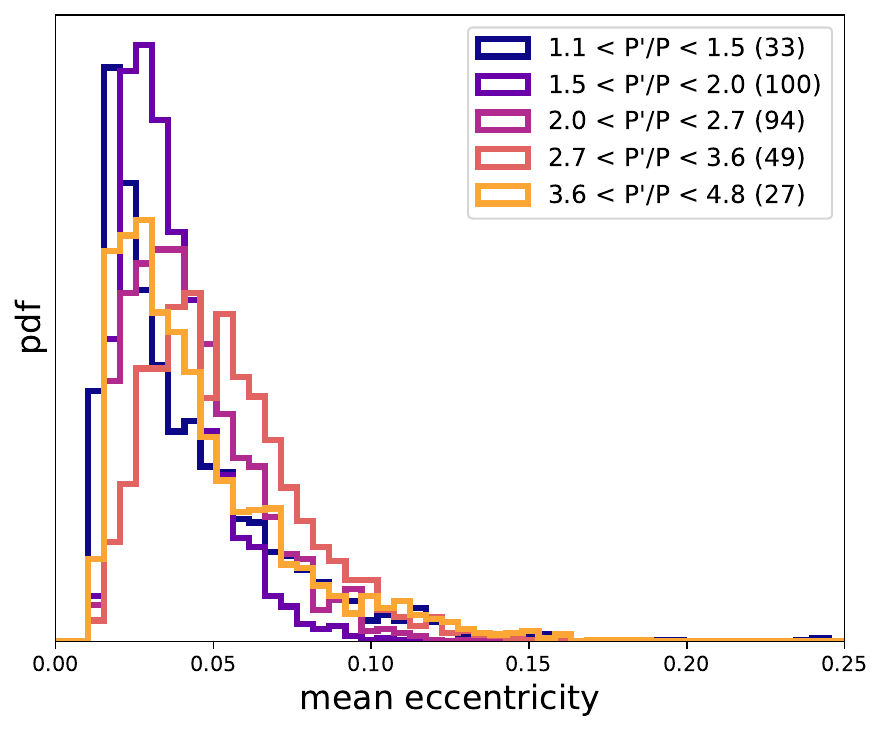}
    \caption{The eccentricity of small planets in two-planet systems as a function of orbital period ratio $P'/P$. The number of planets in each sub-population is indicated in parentheses in the figure legend. No statistically significant difference exists between $\e$ for various sub-populations.}
    \label{fig:ecc-spacing-period-ratio}
\end{figure}

To explore the relationship between eccentricity and orbital spacing, we calculated the period ratio $P'/P$ of planets in two-planet systems, binning the planets in five approximately log-uniform bins between $P'/P = 1.1 - 4.8$. We found no statistically significant differences in eccentricity distribution of the various period ratio bins (Figure \ref{fig:ecc-spacing-period-ratio}), with a typical mean eccentricity $\e \approx 0.04\pm0.02$ in each of the $P'/P$ bins.

\subsection{High multiplicity systems: gap complexity}

To explore the relationship between eccentricity and orbital spacing in higher multiplicity, we calculated the gap complexity $\mathcal{C}$ \citep{GilbertFabrycky2020}) of systems with three or more planets, binning the planets in four approximately log-uniform bins between $0 < \mathcal{C} < 1$. Lower gap complexities correspond to more regular spacings, with $\mathcal{C}=0$ indicating perfect uniform spacing in log-period, $\mathcal{C}=1$ indicating maximum possible spacing disorder. A gap complexity $\mathcal{C} \approx 0.3$ corresponds to a critical value where a system has room for an additional ``missing'' planet to be placed in an over-sized gap between transiting planets. 

\begin{figure}
    \centering
    \includegraphics[width=0.95\linewidth]{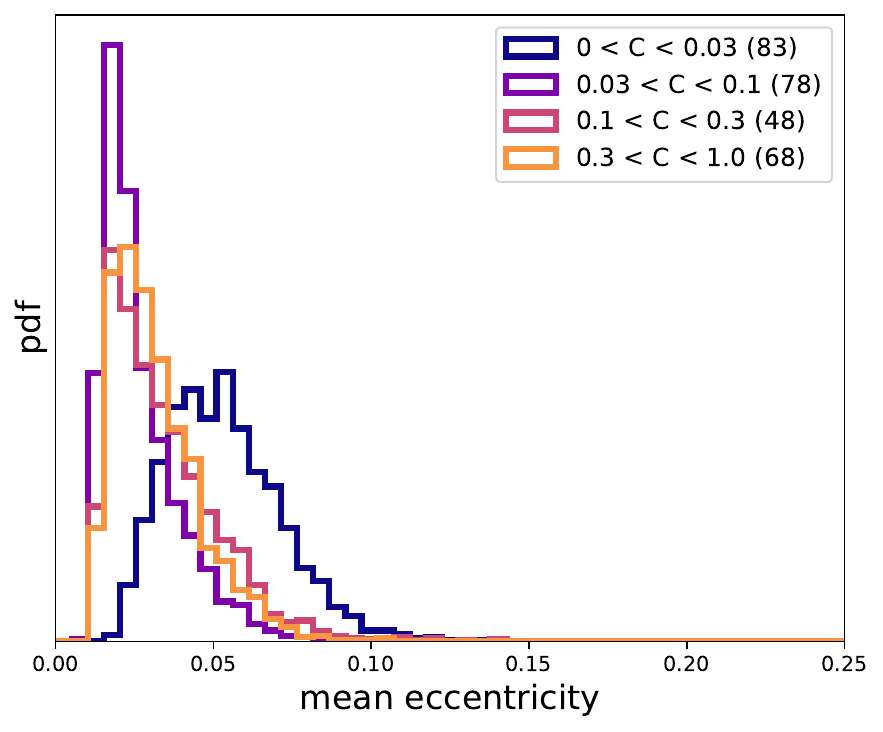}
    \caption{The eccentricity of small planets in high multiplicity ($N \geq 3)$ systems as a function of gap complexity $\mathcal{C}$. The number of planets in each sub-population is indicated in parentheses in the figure legend. No statistically significant difference exists between $\e$ and $\mathcal{C}$ for various sub-populations.}
    \label{fig:ecc-spacing-gap-complexity}
\end{figure}

We found no statistically significant relationships between gap complexity and eccentricity (Figure \ref{fig:ecc-spacing-gap-complexity}). For each of the three bins with $\mathcal{C} > 0.03$, we measured a typical mean eccentricity $\e \approx 0.03^{+0.02}_{-0.01}$. Surprisingly, the most regularly ordered systems ($\mathcal{C} < 0.03$) have elevated eccentricity $\e \approx 0.05 \pm 0.02$, but this difference is consistent within $1\sigma$.

\section{Size architecture}\label{sec:size}

\subsection{Size dispersion}

To explore the relationship between eccentricity and intra-system size inequality (i.e. relative planet size within a system), we calculated the Gini coefficient \citep{Gini1912} on planet radii $G(R_p)$. The Gini coefficient is a measure of statistical dispersion commonly used in studies of economic inequality. A value $G(R_p)=0$ indicates no dispersion (i.e. all planets have equal radii), whereas a value $G(R_p)=1$ indicates maximum dispersion (i.e. one large planet and the rest infinitesimally small). All systems with $N \geq 2$ planets were treated equally in this portion of the analysis. To correct for small $N$, we re-normalized the Gini coefficient as $G' = G \times N / (N-1)$ \citep{Deltas2003}. We binned the planets in five sub-populations such that each bin contained roughly an equal number of planets. 

We found that four out of five bins had statistically equivalent eccentricities, with $\e \approx 0.03^{+0.02}_{-0.01}$. The fifth bin, which had moderate size inequality $0.05 < G(R_p) < 0.10$, indicated slightly elevated eccentricity $\e = 0.059^{+0.017}_{-0.015}$, but after applying the Bonferroni correction for multiple hypothesis testing \citep{ChenFengYi2017}, the significance of this difference falls below $1\sigma$. We searched for unusual characteristics of the anomalously high-$\e$, but we were unable to identify any obvious differences between these systems and the bulk of the population.

\begin{figure}
    \centering
    \includegraphics[width=0.95\linewidth]{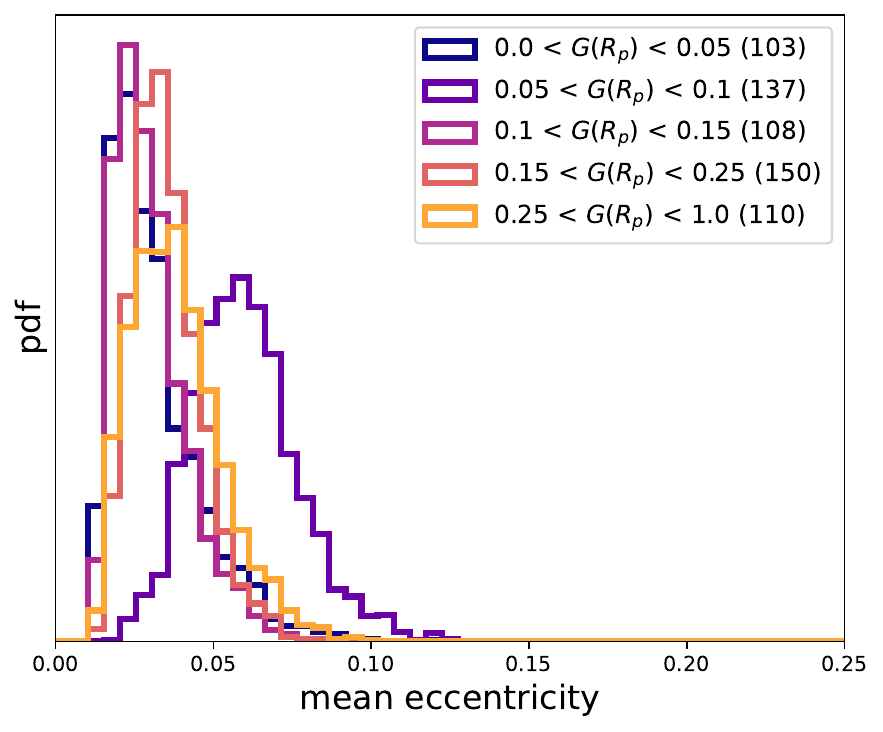}
    \caption{The eccentricity of small planets as a function of intra-system size dispersion, quantified as radius Gini coefficient $G(\Rp)$, re-normalized to account for small number statistics \citep{Deltas2003}. The number of planets in each sub-population is indicated in parentheses in the figure legend. No statistically significant difference exists between $\e$ and $G(\Rp)$ for various sub-populations.}
    \label{fig:ecc-size-gini}
\end{figure}

\subsection{Size ordering}

To explore the relationship between eccentricity and size ordering, we calculated the monotonicity $\mathcal{M}(R_p)$ \citep{GilbertFabrycky2020} of systems with two or more planets, binning the planets in four sub-populations such that each bin contained roughly an equal number of planets. Positive values of $\mathcal{M}$ correspond to systems preferentially arranged with larger planets exterior to smaller planets, whereas negative values correspond to systems preferentially arrange with larger planets interior to smaller planets. $\mathcal{M}$ varies between -1 to +1, with $|\mathcal{M}| = 1$ indicating perfect size ordering and values near $\mathcal{M}=0$ indicating weak or no size ordering, which may arise either from randomly ordered planets or from all planets having nearly the same size.

\begin{figure}
    \centering
    \includegraphics[width=0.95\linewidth]{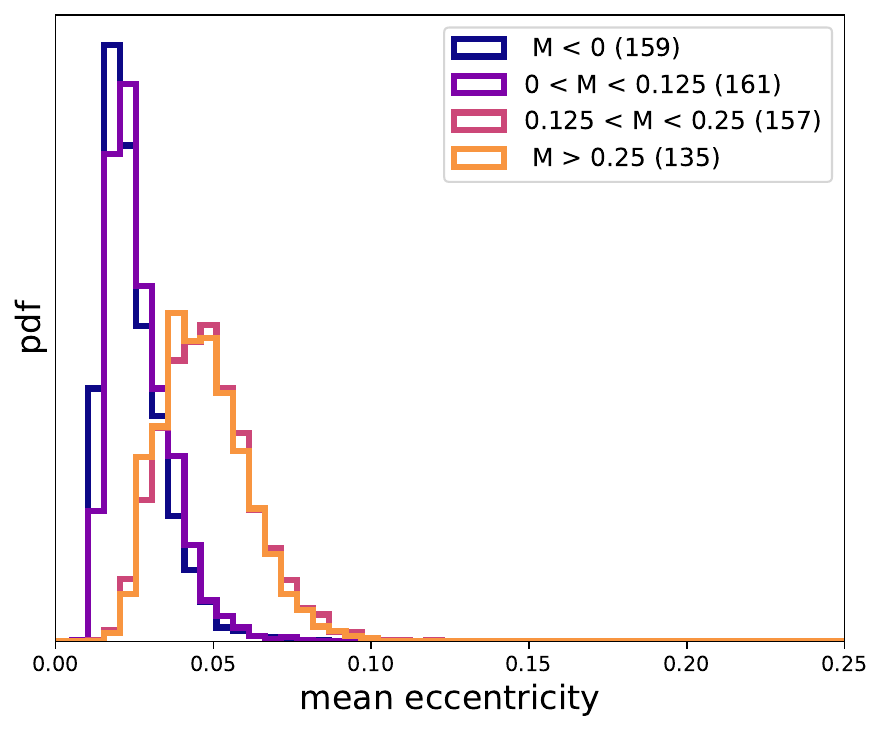}
    \caption{The eccentricity of small planets as a function of monotonicity $\mathcal{M}$. The number of planets in each sub-population is indicated in parentheses in the figure legend. Positive monotonic systems show tentative but sub-significant evidence of elevated $\e$ compared to negatively monotonic or nearly uniform sized ($\mathcal{M}\approx0$) systems.}
    \label{fig:monotonicity}
\end{figure}

We found no statistically significant difference in mean eccentricity as a function of monotonicity (Figure \ref{fig:monotonicity}), although intriguingly the two bins with high positive monotonicity $\mathcal{M} > 0.125$ had somewhat higher --- though not statistically significant --- $\e$ compared to bins with near-zero or negative $\mathcal{M}$.

\section{Dynamical Temperature}\label{sec:dyntemp}

To explore the relationship between mean eccentricity and overall system dynamical temperature, we calculated the ``flatness'' quantity $f$ introduced by \citet{GilbertFabrycky2020}, which captures the extent to which a system deviates from a circular, coplanar architecture. A value $f=0$ corresponds to all planets having $e=0$ and $\Delta i = 0$, whereas $f \rightarrow 1$ corresponds to systems with maximal $e$ and $\Delta i$. We binned the planets in five sub-populations such that each bin contained roughly an equal number of planets.

\begin{figure}
    \centering
    \includegraphics[width=0.95\linewidth]{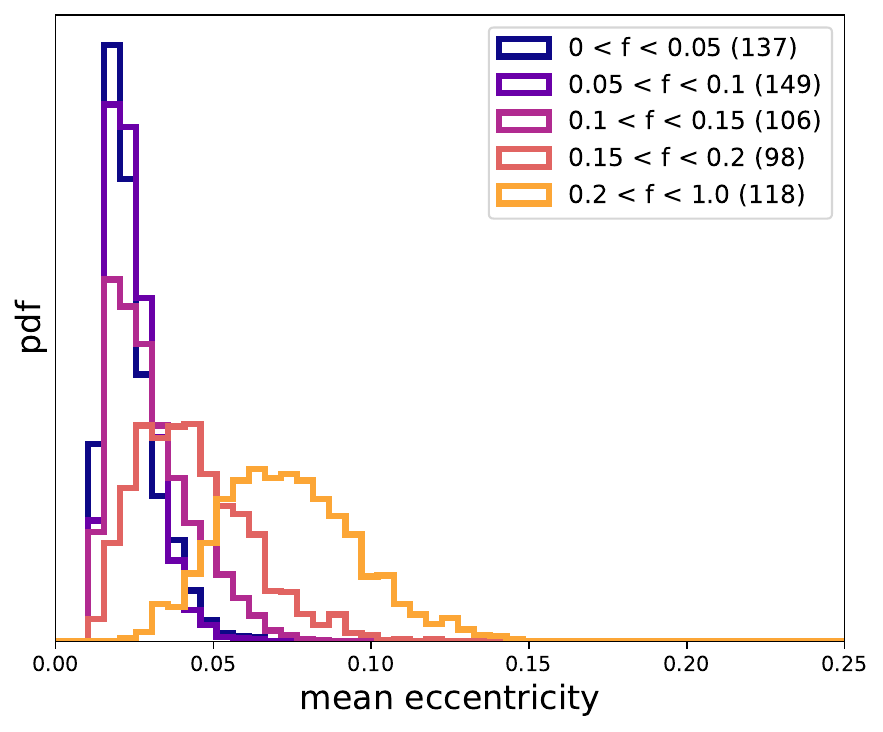}
    \caption{The eccentricity of small planets as a function of overall system dynamical temperature $f$. The number of planets in each sub-population is indicated in parentheses in the figure legend. As expected, mean eccentricity rises with increasing dynamical temperature.}
    \label{fig:ecc-dyntemp}
\end{figure}

We found that mean eccentricity rose with increasing dynamical temperature (Figure \ref{fig:ecc-dyntemp}). Mean eccentricity for the dynamically coolest bin ($f < 0.05$) was $\e = 0.022^{+0.010}_{-0.005}$ and for the dynamically hottest bin ($f > 0.2$) was $\e = 0.072^{+0.023}_{-0.019}$. 

This correlation between $e$ and $f$ is unsurprising, given that $f$ is sensitive to both $e$ and $\Delta i$. Nevertheless, because $f$ is more sensitive to small changes in $\Delta i$ than it is to small changes in $e$, the $e-f$ correlation does imply some coupling between $e$ and $\Delta i$. See \S \ref{sec:discussion} for further discussion of the dynamical implications of these results for joint evolution of eccentricity and inclination.

\section{Discussion}\label{sec:discussion}

We detected several patterns in the relationships between eccentricity and other architectural properties of Kepler systems.

\begin{enumerate}
    \item Small planets inside $P<4$ days show evidence of circularization, whereas those at longer periods do not.
    \item Small planets in multiplanet systems have low $\e \approx 0.03 \pm 0.01$ regardless of whether a giant planet is present in the system or not.
    \item Mean eccentricity is higher by a factor of two in single-transiting systems; 2-planet systems have somewhat higher mean eccentricity than systems with 3+ planets.
    \item There are no strong associations between eccentricity and size or spacing architecture, although some hints of trends exist. Systems with $\mathcal{M} > 0.125$ have slightly elevated $\e=0.05\pm0.02$ compared to systems with $\mathcal{M} < 0.125$, which have $\e=0.02\pm0.01$. The most regularly spaced systems with $\mathcal{C} < 0.03$ also have slightly elevated eccentricities $\e = 0.05 \pm 0.02$.
\end{enumerate}

Below, we discuss several features of the Kepler planet population implied by these observations.

\subsection{Small planets on short orbits are circularized}

Near-zero eccentricity for short-period planets is expected based on predicted tidal circularization timescales \citep{Rasio1996, Jackson2008}. Previous work has clearly demonstrated that giant planets on short orbits tend to be have low eccentricity \citep[e.g.,][]{WinnFabrycky2015}, but tidal circularization has so far only been conclusively demonstrated for a handful of smaller objects. The measurements we have presented here thus constitute the strongest empirical evidence yet that short-period planets across a range of sizes (and presumably densities and compositions) circularize within the lifetime of their host systems.

\subsection{Why are there no strong relationships between eccentricity and architecture in compact multis?}

The lack of any strong relationships between eccentricity and other architectural tracers within the compact multi sample is surprising, as dynamical interactions which excite eccentricity are also expected to excite inclinations \citep[e.g.,][]{Laskar1997, Chatterjee2008, Tomayo2020} and disrupt system regularity \citep{PuWu2015, Izidoro2017, Lambrechts2019}. Thus, before performing this analysis we expected to find as positive correlations between eccentricity and spacing scale $P'/P$, gap complexity $\mathcal{C}$, and size inequality $G(R_p)$. In reality, we detected no such associations.

The apparent lack of a relationship between small planet eccentricity and the presence of a giant planet is even more perplexing because eccentric giant planets are expected to dynamically excite their smaller siblings. To this end, \citet{HeWeiss2023} recently found that compact multis with Doppler-detected outer giants have higher typical gap complexities compared to systems without outer giants. They interpret this result as evidence that giant planets may excite inclinations, producing a population of non-transiting planets that appear as ``missing'' planets in the inner system. Another interpretation is that the outer giants influenced the formation of the inner system. N-body simulations have not yet reached a consensus on the correct explanation for the gap-giant association. \citep{LammersWinn2025, LiveseyBecker2025}.

As a possible resolution to these tensions, we note that only 8 out of 44 giant planets in our present study  have $R_p \geq 6\Re$, which corresponds to $M_p \gtrsim 30 \Me$, based on a mass-radius relationship \citep{ChenKipping2017}. In contrast, 24 out of 27 giant planets studied by \citet{HeWeiss2023} had $M_p \geq 30 \Me$. Furthermore, whereas the transiting giant planets considered here orbit within $a \lesssim 0.5$ au in close proximity to their smaller siblings, the non-transiting giants from \citet{HeWeiss2023} mostly orbit between $a \sim 1{-}10$ au. So, the two populations are qualitatively distinct in terms of both mass regime and architectural context and therefore may represent different formation channels.

We hypothesize that the transiting giants (this study) formed quiescently like ``peas-in-a-pod'' and just barely managed to enter the runaway accretion stage of planet formation \citep{Pollack1996} before the gas disk dissipated, whereas the larger non-transiting giants \citep{HeWeiss2023} experienced substantially more dynamical evolution. Indeed, \citet{Lee2019} suggests that hydrodynamic flows can naturally stall runaway accretion for planets with solid cores $\lesssim10\Me$ which might prevent embedded ``giant peas'' from accreting sufficient mass to dynamically disrupt their systems. If this formation picture is correct, then these giant-hosting systems must have formed while the gas disk was still present, which would have damped down eccentricities and inclinations.

Still, giant planets --- even undersized ones --- should dynamically excite their siblings at least a little, leaving open the question of why we do not observe any obvious relationships between $\e$ and $\mathcal{C}$. A possible resolution is that the strong geometric detection biases of the transit method make observation of high-$\Delta i$, and consequently high-$e$, multiplanet systems unlikely. If indeed mutual gravitational interactions between planets dominate the dynamical evolution of planetary systems, eccentricity and inclination are expected to be tightly correlated \citep[e.g.,][]{Laskar1997, Chatterjee2008, Tomayo2020}, a theoretical expectation which has been borne out observationally across a wide range of celestial objects \citep{Xie2016} and in population synthesis \citep{He2020}. Empirically $e\approx2i$ (when $i$ is expressed in radians); this relationship arises naturally due to exchange of angular momentum between the one degree of freedom for inclination and two degrees of freedom for eccentricity.

From this scaling relation, when $e=0.1$, we may expect $\Delta i\approx2^\circ$. For a typical \textit{Kepler} planet ($P = 20$ days) orbiting a sun-like star, an inclination $\Delta i = 2^\circ$ relative to a center-crossing orbit with impact parameter $b=0$ is sufficient to tilt the planet so that $b > 1$ and the planet is no longer seen to transit. Consequently, in multi-planet systems any $e \gtrsim 0.1$ is likely to lead to $\Delta i \gtrsim 2^\circ$, making the observation of multiple transiting planets unlikely. Although non-zero eccentricity enhances transit probability \citep{Barnes2007, Burke2008}, the enhancement scales only as $(1-e^2)^{-1}$, so the effect is negligible for low eccentricities. So, in the absence of a lucky coincidental alignment, in order to detect multiple planets, we require $\Delta i \lesssim 2^\circ$, which implies $e \lesssim 0.1$. These values agree with empirically measured values for $\Delta i \approx 2^\circ$ \citep{Fabrycky2014} and $\e \approx 0.03$ (this work) in Kepler compact multis.

The above reasoning leads us to conclude that the data are inconsistent with astrophysical processes which excite eccentricities without also inciting inclination, providing some support for planet-planet interactions as the primary source of dynamical sculpting within systems. An alternate interpretation is that Kepler compact multis experienced little dynamical excitation after settling into their final arrangements of masses and periods following gas disk dispersal.

\subsection{Kepler singles are dynamically heated and have low intrinsic multiplicity}

The observed eccentricity enhancement for single-transiting vs. multi-transiting systems has been well-established by previous analyses \citep{VanEylen2019, Gilbert2025}. Here, we refine this trend and show that systems with three or more transiting planets have lower $\e$ compared to systems with two transiting planets. Previous studies had identified a tentative anti-correlation between multiplicity and eccentricity for planets detected by both transits and radial velocities \citep{Xie2016, ZinziTurrini2017, Zhu2018}. This anti-correlation is also expected on theoretical grounds. For example, the ``maximum AMD'' model developed by \citet{He2020} automatically produces lower eccentricities in systems with more planets when the angular momentum deficit \citep[AMD;][]{Laskar1997, LaskarPetit2017} is evenly apportioned to each planet per unit mass.

\begin{figure}
    \centering
    \includegraphics[width=0.95\linewidth]{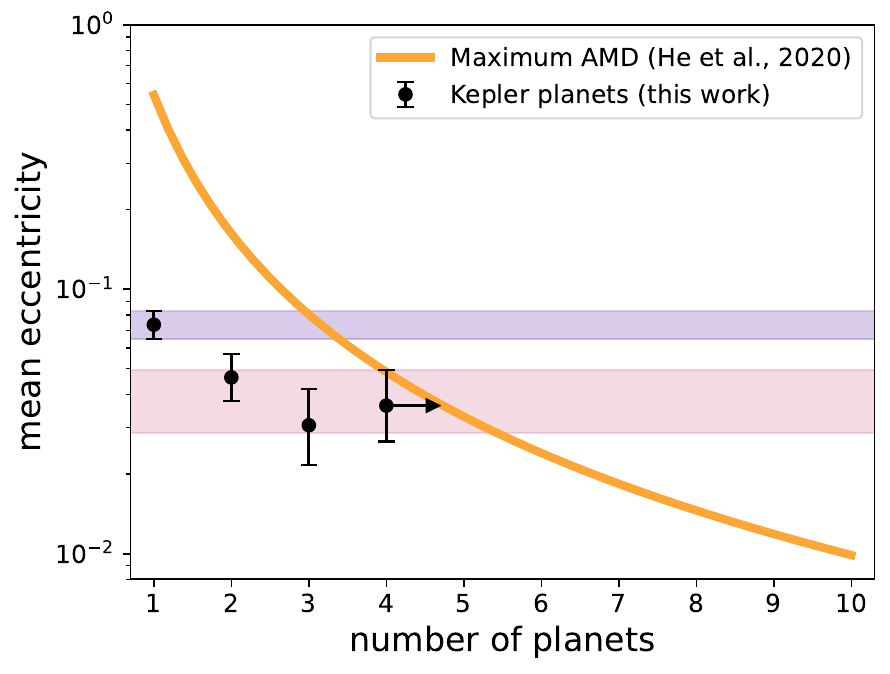}
    \caption{Mean eccentricity as a function of number of planets, comparing our empirical measurements (black points) to population synthesis predictions from the ``Maximum AMD'' model of \citet{He2020} (gold line). The purple and pink shaded are drawn to aid the eye in assessing the credible range of eccentricities for single-transiting and multi-transiting systems, respectively. Our empirical measurements are based on observed multiplicity, whereas the \citet{He2020} theoretical values are based on intrinsic multiplicity. There exists a clear tension between measurement and prediction which can be resolved if we assume that the true multiplicity of Kepler single-transiting systems is ${\sim}3$, while the true multiplicity of Kepler multi-transiting systems is ${\sim}4{-}6$.}
    \label{fig:max_amd}
\end{figure}

In Figure \ref{fig:max_amd}, we plot our empirical $\e$ values as a function of the number of planets compared to the theoretical $\e{-}N$ relationship predicted by the maximum AMD model. This comparison is not exactly one-to-one, as maximum AMD model describes eccentricity as a function of \textit{intrinsic} multiplicity, whereas we measure eccentricity as a function of \textit{observed} multiplicity, which is subject to strong selection effects and biases. From inspection of Figure \ref{fig:max_amd}, it appears that the tension between prediction and observation can be resolved if single-transiting systems usually have intrinsic multiplicity $N \sim 3$ and multi-transiting systems usually have intrinsic multiplicities $N \sim 4{-}6$.

\section{Summary \& Conclusion}\label{sec:conclusion}

In this study, we analyzed the eccentricities of \numplanets transiting planets from the Kepler census, exploring the relationship between eccentricity, transit multiplicity, and system size/spacing architecture. We found that eccentricity and transit multiplicity are anti-correlated but detected no strong associations between eccentricity and size or spacing architecture. We also demonstrated that small planets on short orbits ($P < 4$ days) show evidence of tidal circularization \citep{Rasio1996, Jackson2008}. Within the sample of compact multis, we found that only the most dynamically heated systems host planets with unambiguously elevated eccentricities. We interpret these finding as evidence in favor of either a quiescent formation history or as evidence against dynamical processes which excite eccentricity without also exciting inclination.

Our non-detection of strong associations between eccentricity and system architecture properties should not, however, be taken as conclusive evidence that no trends exist. Indeed, we observed some hints of associations between eccentricity and system size architecture, with positively monotonically ordered systems exhibiting somewhat elevated eccentricities compared to their non-ordered or ``inverted'' counterparts. We also found that the most regularly spaced systems also exhibit elevated eccentricities. It may yet be possible to identify additional relationships in the data by adapting our one-dimensional hierarchical slicing method into a more sophisticated multi-factor model capable of measuring more subtle astrophysical patterns.

\section*{Acknowledgments}
We thank Dan Fabrycky and Lauren Weiss for insightful conversations.

All \textit{Kepler} lightcurves used in this analysis are hosted by the Mikulski Archive for Space Telescopes (MAST) at \citet{10.17909/t9488n}. Stellar parameters are available from the Gaia-Kepler catalog \citep{Berger2020}.

Funding for this work was provided by a University of California, Los Angeles set-up award to E.A.P. and by the Heising-Simons Foundation Award \#~2022-3833

\software{\texttt{astropy} \citep{astropy:2013, astropy:2018, astropy:2022},
          \texttt{numpy} \citep{numpy:2020}, 
          \texttt{scipy} \citep{scipy:2020}, 
          \texttt{pymc3} \citep{pymc3:2016}
          }

\clearpage

\bibliography{main}{}
\bibliographystyle{aasjournalv7}

\end{document}